\documentclass[conference]{IEEEtran}
\IEEEoverridecommandlockouts
\usepackage{float}  
\usepackage{tabularx} 
\usepackage{cite}
\usepackage{amsmath,amssymb,amsfonts}
\usepackage{algorithmic}
\usepackage{graphicx}
\usepackage{tcolorbox}
\usepackage{caption,subcaption}
\usepackage{booktabs}
\usepackage{adjustbox}
\usepackage[table,xcdraw]{xcolor}
\usepackage{textcomp}
\usepackage{multirow}
\usepackage{multicol}
\usepackage{booktabs}
\usepackage{bigstrut}
\usepackage{verbatim}
\usepackage{array}
\usepackage{rotating}
\usepackage{dblfloatfix}
\usepackage{soul}
\usepackage[a4paper, total={184mm,239mm}]{geometry}

\begin{document}

\title{Generalizable Verilog Modeling Framework for Synchronous and Asynchronous Superconducting Pulse-Based Logic Gates}

\author{Elisabeth Feng, Robert S. Aviles, Peter A. Beerel,~\IEEEmembership{Senior Member, IEEE} \thanks{This work has been supported by ARL DEVCOM under the FSDL: ColdPhase project, grant number W911NF-24-1-0317.

The authors are with the Department of Electrical and Computer Engineering,
University of Southern California, Los Angeles, CA 90007 USA (e-mail: fenge@usc.edu;
rsaviles@usc.edu; pabeerel@usc.edu)}}

\maketitle

\begin{abstract}
Superconducting Single Flux Quantum (SFQ) logic offers a promising platform for ultra-low-power, high-frequency computing. However, their pulse-based nature poses challenges for scalable modeling, design, and verification using conventional hardware description languages (HDLs), which are designed for level-based digital logic. Prior efforts have required complex Verilog support modules to enable Standard Delay Format (SDF) compatibility and have provided limited coverage of SFQ cell types. This work presents a Verilog-based modeling framework for SFQ gates that enables functional and timing verification while maintaining compatibility with Standard Delay Format (SDF) back annotation and is the first framework to support both synchronous and asynchronous SFQ gates. The proposed models are validated through device-level simulations, demonstrating correct functionality and timing constraint coverage. RTL simulation of mixed synchronous–asynchronous circuits further demonstrate the utility of the proposed framework.
\end{abstract}


\section{Introduction}

As CMOS devices approach fundamental power and scaling limits \cite{Moores}, superconductor electronics (SCE) has emerged as a promising alternative for ultra-low power, high-frequency computing.  Superconductor integrated circuits span 32GHz microprocessors \cite{32GHz_Processor}, digital quantum computing controllers \cite{Quantum_controller}, %cryptographic accelerators \cite{SCE_NTT}
and space-based optical sensors \cite{Gao2025}. 

\begin{figure}[htbp]
  \includegraphics[width=1\linewidth]{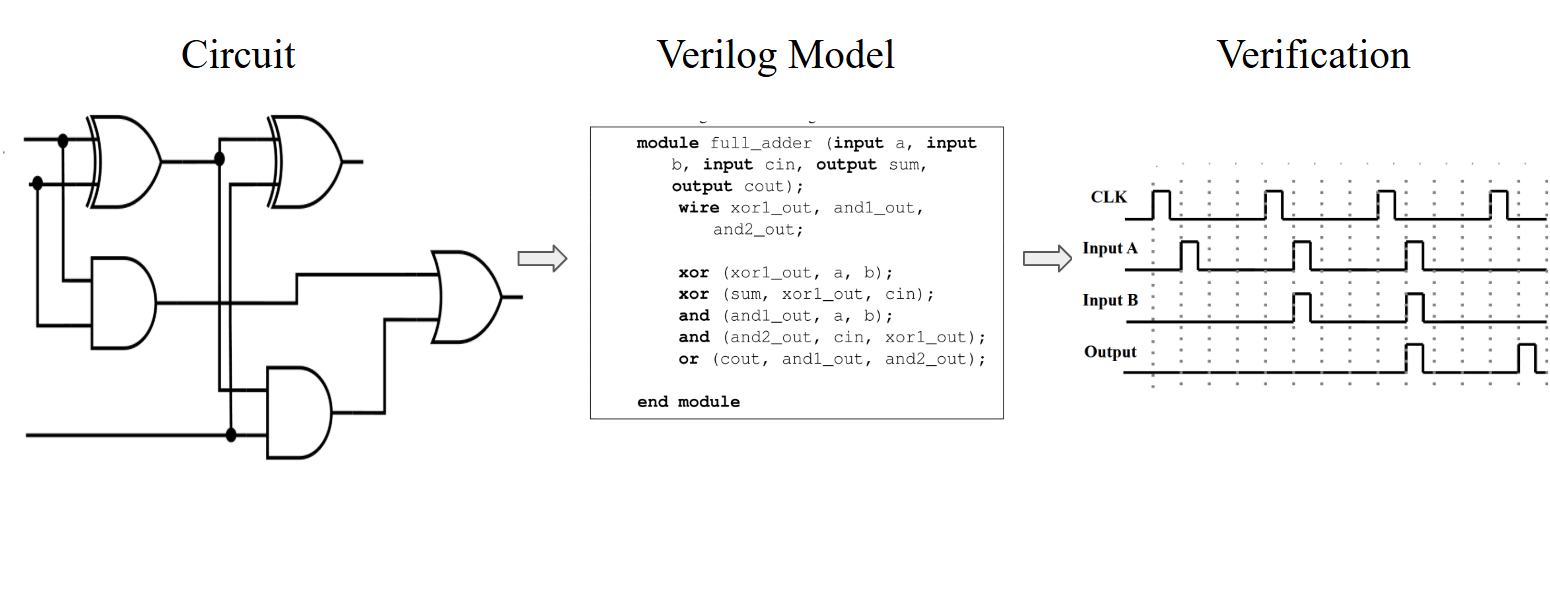}
  \vspace{-0.6cm}
\centering
  \caption{Verilog-based modeling, design, and verification flows}\label{fig:RTLFlow}
\end{figure}

Despite significant progress, realizing large-scale SCE systems remains challenging due to fabrication constraints and electronic design automation (EDA) challenges arising from the pulse-based nature of superconductor devices such as Single Flux Quantum (SFQ) logic \cite{Nb_Fabrication,SFQ_EDA_2024}. Full-custom manual design based on circuit-level analog simulators, including JoSIM \cite{Josim} and PSCAN \cite{PSCAN}, accurately model superconductor behavior but quickly become computationally prohibitive for large circuits. More recent efforts have focused on developing an ASIC-based approach with a fully characterized cell library \cite{qPALACE}. Here, Verilog models of the SFQ cell library that capture their timing constraints and support post-routing delays is essential for modeling, design, and functional verification at scale as shown in Fig.~\ref{fig:RTLFlow}. More specifically, to be consistent with modern Verilog flows, it is preferable for the timing constraints to be modeled using specify blocks and the post-route timing to be captured with 
Standard Delay Format (SDF) back-annotation flows. 

However, Verilog and SDF were both designed for level-based digital logic and do not naturally support the pulse-based semantics of SFQ circuits. A particular challenge arises with asynchronous SFQ gates \cite{First_Async}, 
which exhibit a temporal pulse retention behavior that has not been supported in prior modeling efforts~\cite{structural_modeling, First_Verilog_RSFQ, SFQ_HDL_2003, VHDL_RSFQ, SystemVerilog_2020}.  Furthermore, previous methods to integrate SDF support have relied on complex semantics and multiple helper modules~\cite{SystemVerilog_2020}.

To address this limitation, this paper proposed a generalizable Verilog-based framework capable of modeling both synchronous and asynchronous SFQ gates while maintaining compatibility with SDF back annotation. In particular, this work makes the following contributions:
\begin{itemize}
    \item \textbf{Simplified Gate Modeling Framework}: A modeling framework that reduces Verilog module complexity by relocating hierarchical timing references from Verilog descriptions to the SDF file, eliminating the need for auxiliary interconnect modules.
    \item \textbf{Asynchronous SFQ Gate Abstraction}: The first Verilog abstraction supporting asynchronous SFQ gates, introducing a novel methodology for representing pulse retention windows using SDF timing constructs.
    \item \textbf{Mixed-Mode SFQ Verification}: Demonstration of the benefits of combining synchronous and asynchronous SFQ gates, including a four-bit T1-cell based multiplier, validating correct functional behavior within a unified HDL-based environment.  
\end{itemize}

The remainder of this paper is organized as follows. Section II discusses modeling challenges for synchronous and asynchronous SFQ gates and reviews limitations of prior approaches. Section III introduces the proposed modeling framework. Section IV presents verification against circuit-level simulations using JoSIM. Finally, Section V concludes the paper and discusses future directions. 

\section{Background}
\subsection{Single-Flux Quantum}

\begin{figure}[htbp]
  \includegraphics[width=1\linewidth]{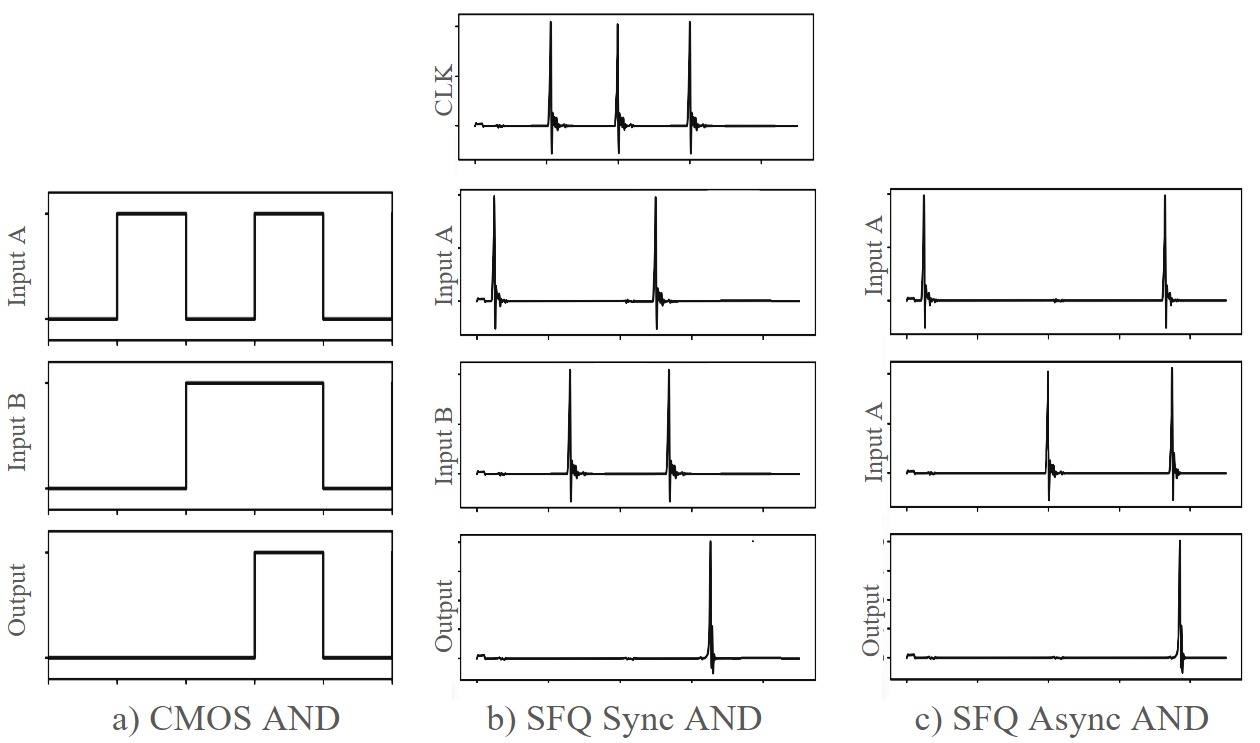}
    \vspace{-0.6cm}
\centering
  \caption{Functional comparison between CMOS AND, SFQ sync AND, and SFQ async AND gates}\label{fig:CMOS_comp}
\end{figure}

Unlike CMOS circuits, which represent information using voltage levels, Single-Flux Quantum (SFQ) logic encodes information through the presence or absence of a magnetic flux quantum as shown in Fig.~\ref{fig:CMOS_comp}. Josephson Junctions (JJs) produce a discrete SFQ voltage pulse of millivolt amplitude that lasts only a few picoseconds, serving as the marker for flux presence \cite{RSFQ}. Most conventional SFQ logic cells operate synchronously, evaluating stored inputs upon the arrival of a clock pulse as shown in Fig.~\ref{fig:CMOS_comp}b. During operation, the generation of an SFQ pulse typically destroys the stored flux within the cell, leading to a destructive read-out (DRO) behavior. However, internal feedback mechanisms can be incorporated to enable flux storage, allowing nondestructive read-out (NDRO) operation.
The gate-level clocking used in synchronous SFQ circuits results in deeply pipelined logic structures. To maintain correct functionality, typically all logic paths are fully balanced, requiring significant buffer insertion to equalize path delays. To address this overhead, optimization strategies have been recently proposed that schedule clock arrival times across the circuit to relax path-balancing constraints~\cite{AvilesMultiPhase,DelayBalancing,SFQ_Optimizing}. 
To further reduce area overhead, several approaches integrate asynchronous gates to implement clock-less combinational logic~\cite{Async_adder,CompoundGates}. In these designs, arriving pulses are temporarily held within superconducting loops before dissipating, introducing a finite \textit{retention window} that defines timing windows for pulse interactions~\cite{Async}. For coincidence-based gates, such as an asynchronous AND gate, output pulse generation is not driven by a clock but instead occurs when both input pulses arrive within overlapping retention windows, as in Fig.~\ref{fig:CMOS_comp}c. In contrast, for OR gates, only a single input pulse need arrive to generate an output pulse. Importantly, overlapping input retention windows still produce only a single output pulse.

Hybrid synchronous–asynchronous cells can further improve circuit density. The T1 cell \cite{T1_first}, uses a single input line capable of receiving multiple SFQ pulses within a single clock cycle, with each pulse toggling the stored internal flux state. An output pulse is generated whenever the internal state returns to the zero-flux condition. This output pulse occurs on an asynchronous output port when triggered by an input pulse or upon the synchronous output port when triggered by the arrival of a clock pulse. The T1 cell can implement a one-bit full adder, where the synchronous port corresponds to the sum output and the asynchronous port corresponds to the carry output. This implementation requires approximately $40\%$ of the area of conventional SFQ adders due to the reduced number of Josephson junctions~\cite{T1_cell}. %Such designs demonstrate how nonconventional SFQ gates can significantly improve circuit density and integration. 

\subsection{ASIC Design Flows}

In typical ASIC design flows for CMOS design, a high-level behavioral specification is automatically synthesized to a gate-level netlist of characterized cells, both typically described in Verilog~\cite{NanniBook}. The cells are then physically placed and routed and a back-annotation flow computes all interconnect delays as well as updates to the timing of each cell instance, storing the results in Standard Delay Format (SDF). The SDF file in combination of with Verilog netlist enables fast timing accurate simulation that does not relying on lower-level analog models.

More specifically, Verilog \textit{specify} blocks \cite{VerilogRef} define input to output pin delays (such as clk-to-q) as \textit{IOPATH} delays as well as timing checks such as \textit{SETUP} and \textit{HOLD}. SDF back annotation then enables overriding default parameter values with per instance values. Additionally, SDF has an \textit{INTERCONNECT} construct to model net delays between cell instances capturing the wire delay between cells. 

In SFQ circuits, interconnect delays typically represent a majority of the clock period and the ability to employ accurate timing constraint checks is fundamental to ensuring correct functionality. As SFQ technology matures and scales, the ability to adapt such existing design flows that are supported by a wide variety of open-source and commercial tools becomes critical. 

\subsection{Existing Models for SFQ}

This benefit of Verilog modeling of SFQ circuits has been realized since the 
1990s~\cite{structural_modeling,First_Verilog_RSFQ}. 
%The idea of functional modeling is to abstract gate behavior with events capturing the transfer of discrete fluxes. Gates should be fully characterized by timing parameters governing flux movement which typically include setup time, hold time, and propagation delay. 
%
The first proposed Verilog model captured the transferring of discrete voltage pulses between synchronous gates, but did not support standard CMOS design flow such as SDF-based back annotation and timing violation reporting~\cite{First_Verilog_RSFQ}. Other models~\cite{SFQ_HDL_2003} have attempted to model timing constraints, however, lacked the ability to describe interconnection delay between circuit elements, 
relied on hard-coded delay characteristics, 
and did not support asynchronous SFQ cells.
VHDL-based models of SFQ cells have also been presented, addressing conditional timing parameters such as data dependent propagation delay through finite state machines description`\cite{VHDL_RSFQ}. While achieving high timing accuracy, this methodology requires unique timing arcs and checks for each gate that do not support standard SDF-based flows. 

The most recent effort proposes an SDF-compatible SystemVerilog (SV) model using SV interfaces and auxiliary helper modules to separately model gate and interconnect  delay. As \textit{specify} timing blocks 
do not support references to SystemVerilog interfaces, these helper module provided Verilog ports to anchor specified timing paths~\cite{SystemVerilog_2020}. However, these modules add unnecessary complexity to the SDF creation requiring the separation of data delay and timing checks into two cell types within SDF. Additionally, destructive read out characteristic of pulse consumption was inherently tied to the modules, preventing the ability to model non-destructive read out cells (NDRO) 
or extend this framework to asynchronous gates. 

In fact, to the best of our knowledge, there is no proposed Verilog model for asynchronous SFQ gates nor any methodologies to describe their temporary pulse retention using a standard CMOS flow. %This retention window is not a typical timing characteristic meant to be described through SDF back annotation or violation checks. 

\section{Proposed Verilog Modeling Framework}

Our proposed Verilog framework converts transient pulse signals into persistent internal voltage levels to maintain data persistence and perform logic operations while maintaining discrete output pulse generation. 
For synchronous gates, internal data persists until the next clock signal as opposed to asynchronous gates where internal data persistence is instead governed by retention windows that define the time before decay or output triggers. To ensure SDF back-annotation compatibility and scalability, our proposed framework uses generalizable blocks to describes the internal data, output generation, logic functionality, and timing characteristics. In particular, this framework formally characterizes asynchronous 
retention windows as an SDF back-annotated parameter. 

\subsection{Synchronous SFQ Gates}

\begin{figure}[htbp]
  \includegraphics[width=1\linewidth]{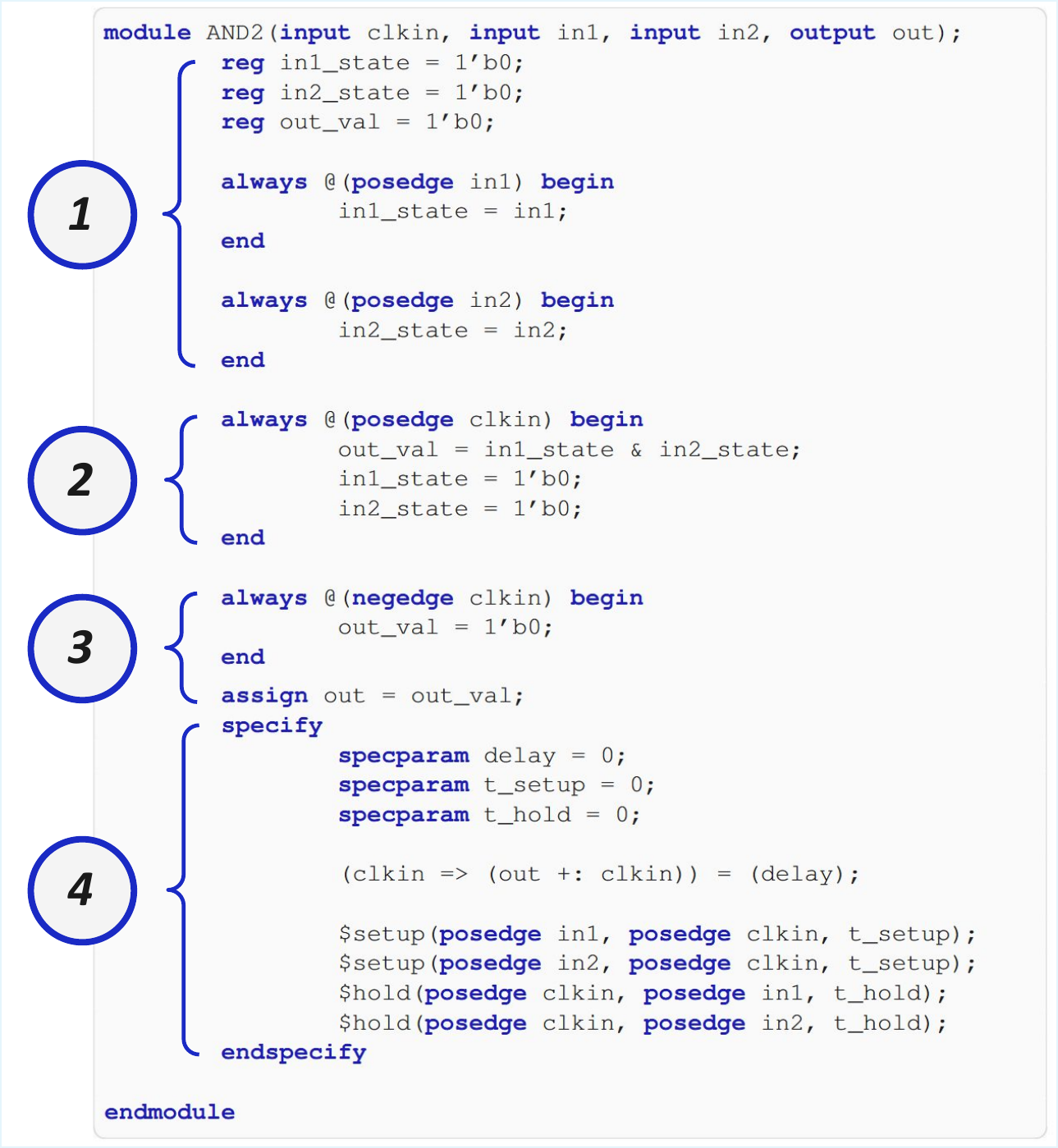}
    \vspace{-0.6cm}
\centering
  \caption{Synchronous AND2 Verilog module}\label{fig:syncAND}
\end{figure}

More specifically, our synchronous SFQ gate model 
is based off of the four blocks shown in Fig.~\ref{fig:syncAND}: 1) Initialization/Input Capture 2) Clock Triggered Logic functionality 3) Output Pulse Generation and 4) Specify Timing Block. 

1) Initializations/Input Capture: Persistent internal voltage levels are achieved through capturing of input pulses in internal state registers (denoted by the suffix \_state) providing a stored representation of pulses. Verilog procedural logic blocks such as \textit{always} blocks are resolved at compile time, only allowing fixed simulation delay specification to impact timing \cite{VerilogRef}. Such fixed delays cannot be overridden via SDF. In contrast, timing along \textit{continuous assignment} paths to output ports can be overridden via SDF \cite{VerilogRef}. Thus, registers within procedural blocks cannot have their data be delayed, but data propagation from registers to outputs can have the data be delayed.  
With this motivation in mind, our model defines a intermediate register to drive the output port 
(denoted by \_val) that is driven by the associated internal state registers. Thus each input port requires a state register while each output port requires a \_val register. 

2) Clock Triggered Logic Functionality: On the rising edge of the clock, the output value register updates accordingly to the defined logic functionality.  To model pulse consumption (destructive read out) upon receiving the clock signal, internal state registers are reset. This reset condition can be omitted to simulate nondestructive read out cells.   

3) Output Pulse Generation: Since the output value register was updated at the arrival of the clock, the output port is continuously assigned to this value. To avoid relying on fixed simulation delay constructs to reset the output register and emulate a discrete pulse width, we propose to tie the pulse duration directly to the clock pulse parameters. Specifically, the falling edge of the clock resets the output value register. This approach eliminates the need for hard-coded simulation delays, and establishes defined clock-to-output (clk-to-Q) relationship. If rising and falling edges of a single output pulse were driven by different inputs, there is no unified causal path for consistent back annotation to apply to, complicating output dependency. Ensuring singular paths provides clear relationships for delays to be applied and ensuring SDF compatibility. 

4) Specify Timing Block: Initialized delay parameters within the \textit{specify} block are overwritten by the SDF through back annotation for each module instantiation. Specific edges for timing checks are defined in relation to each other. 

We also note that simulators, such as ModelSim, by default employ inertial delay to \textit{specify} block timing. Under inertial delay schema, signals whose pulse widths are shorter than the applied delay are suppressed. To circumvent this behavior, simulator specific flags such as $+transport\_path\_delay$ ~\cite{structural_modeling} and $+transport\_int\_delay$~\cite{structural_modeling}  select transport delays to ensure all pulses of width smaller than their delays are still propagated on paths and interconnects. 

Our synchronous model thus supports generalized logic functionality, while ensuring SDF compatibility without additional support modules and without forcing destructive read out. 

\subsection{Asynchronous SFQ Gates}

\begin{figure}[htbp]
  \includegraphics[width=1\linewidth]{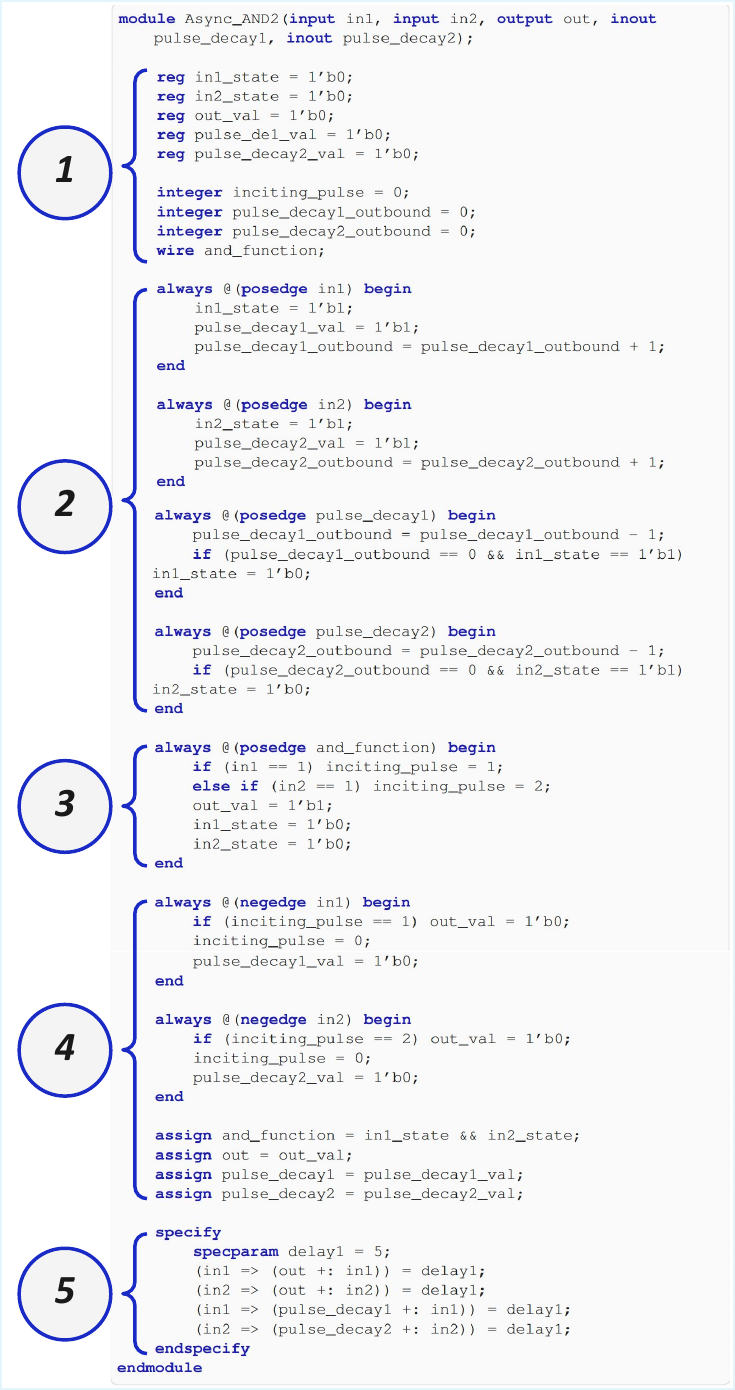}
    \vspace{-0.6cm}
\centering
  \caption{Asynchronous AND2 Verilog module}\label{fig:AsyncAND}
\end{figure}

Similarly, we propose to generalize asynchronous SFQ gates functionality based off of five blocks shown in Fig.~\ref{fig:AsyncAND}: 1) Initialization/Input Capture 2) Pulse Retention Handling 3) Input Trigger Logic Functionality 4) Output Pulse Generation and 5) Specify Timing Block.  While the general principle of pulse width and internal states is similar to the framework for synchronous gates, these gates exhibit additional functionality based on retention windows for pulse storage. The fundamental blocks are shown in Fig.~\ref{fig:AsyncAND}.

1) Initialization/Input Capture: Similar to synchronous gates, each input will have a corresponding internal state register updated at the arrival of an input pulse. However,  modeling pulse retention requires additional internal state registers and counters for each input to track active signals which will be described in the following section.

Because IOPATH delay is no longer synchronized to a single clock edge, the precise relationship between the triggering input stimulus and the output will be tracked with an \textit{inciting\_pulse} variable to preserve width reproduction and timing correlation. 

2) Pulse Retention Handling: To model the expiration of an input pulse retention window and corresponding resetting of internal state registers, we introduce a propagated decay signal mechanism. Upon the arrival of an input pulse, a corresponding signal (pulse\_decay) begins propagating towards an output port. The arrival of this signal to the port serves as a marker, signifying the end of the pulse's physical retention and triggering the reset of internal states. 

The challenge, in the case of an asynchronous AND gate, is that subsequent input pulses can arrive before the previous pulse has effectively decayed which causes the retention window to extend. To prevent premature state termination, we propose to track
the number of outbound decay signals  (pulse\_decay\_outbound) such that only the final arriving signal correctly resets the internal state register.  This behavior is formalized in Table \ref{Table:AsyncGate}.

\begin{table}[H]
\renewcommand{\arraystretch}{1.4}
\begin{tabularx}{\linewidth}{@{}lXX@{}}
  \toprule
  \textbf{Event} & \textbf{Precondition} & \textbf{Effect} \\
  \midrule

  $\mathsf{InputPulseReceived}$
    & ---
    & $\mathit{in\_state} \leftarrow 1$,\newline
      $\mathit{decay\_outbound} \mathrel{+}=$ \\[4pt]
  $\mathsf{OutputGenerated}$
    & ---
    & $\mathit{in\_state} \leftarrow 0$ \\[4pt]
  $\mathsf{DecaySignalReceived}$
    & $\mathit{decay\_outbound} > 0$
    & $\mathit{decay\_outbound} \mathrel{-}=$ \\
  \bottomrule
\end{tabularx}
\caption{
Logic for asynchronous gate retention windows.}
\label{Table:AsyncGate}
\end{table}

To describe this behavior in Verilog, we propose the use of an internal propagating signal so that the retention window can be modeled as an SDF-compatible timing arc. More specifically, to maintain compatibility with our established model, we create an internal pulse\_decay\_state register to represent the signal propagating towards the output value register (pulse\_decay\_val). As a result, additional output ports are declared that do not serve as functional interfaces but enables an \textit{IOPATH} delay that can be controlled via \textit{specify} block. 

3) Input Triggered Logic Functionality: Without a singular procedural block to update the output value as implemented with synchronous gates shown in Fig.~\ref{fig:syncAND} block 2, the logic functionality must be continuously assigned. As Verilog prohibits registers from being driven by continuous assignments, an intermediate wire serves as the logic functionality signal. Upon assignments to this wire, the output value register updated and drives the final output port. This structure allows the model to maintain parameterized \textit{IOPATH} delay. 

Within this framework, as seen in Fig.~\ref{fig:AsyncAND} Block 3, the model monitors the moment the output value is launched to identify the inciting input pulse. Storing the inciting input pulse maintains the casual input to output relationship that is necessary for the \textit{specify} block timing. 

4) Output Pulse Generation: Without a global clock to determine standardized pulse width, output generation is tied to the inciting input pulse through the previously defined tag. Input pulse width generation will now determine the system's parameters and be held consistently across gates. 

5) Specify Timing Block: The specify timing parameters are handled identically to synchronous gates. The propagation delay signal used for the retention window will be treated in the same fashion as \textit{IOPATH} delays. Similar to synchronous, simulator level flags may need to be applied.

\subsection{T1 Cell Modeling}
\begin{figure}[htbp]
  \includegraphics[width=0.9\linewidth]{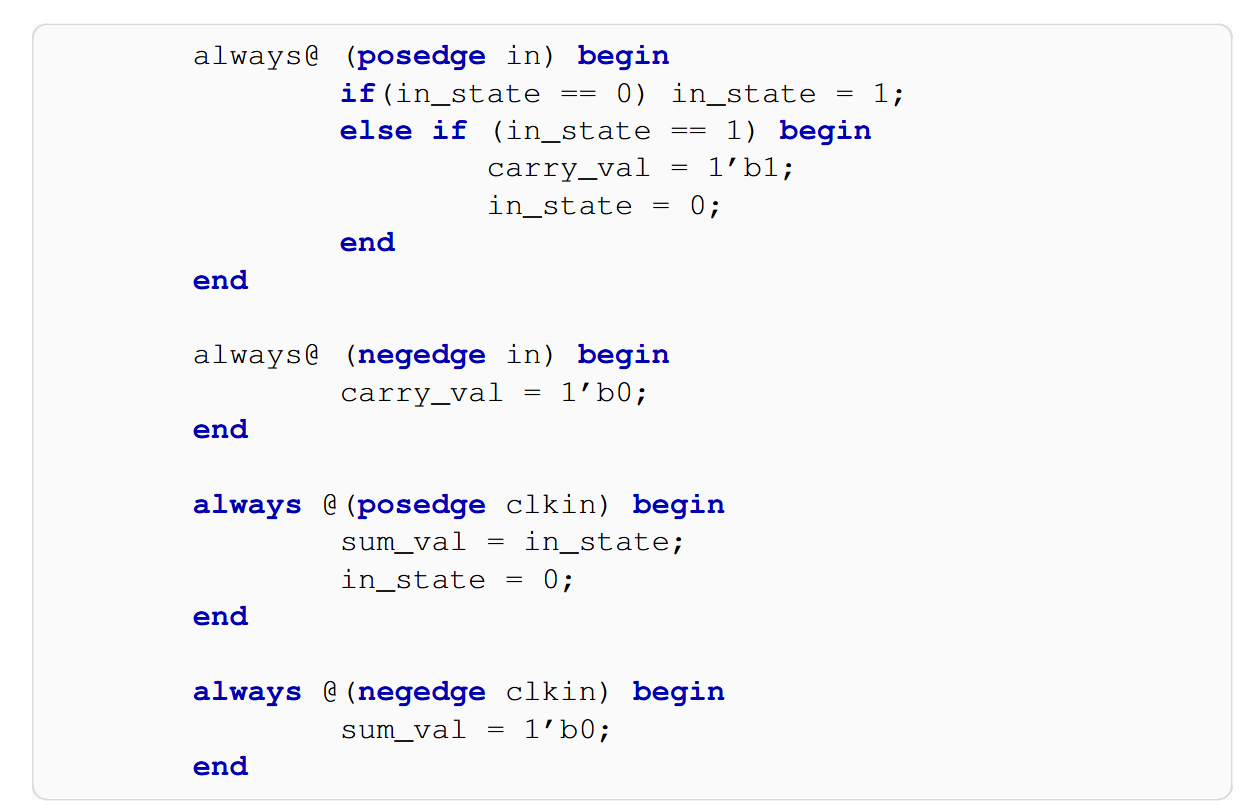}
    \vspace{-0.2cm}
\centering
  \caption{T1 synchronous and asynchronous output handling}\label{fig:T1Veri}
\end{figure}

As T1 cells exhibit hybrid synchronous and asynchronous behavior, our model reflects this by combining elements of both frameworks. Sum is treated as an synchronous output, tied directly to the clkin pulse. Carry, however, borrows the concept of input triggered functionality requiring a direct relation with the single input pulse. Since there is only a single input line, a retention window is not necessary for enforcing decay signal behaviors. The following snippet in Fig.~\ref{fig:T1Veri} shows the input triggered logic functionality of carry and the clkin triggered logic functionality of sum is shown in.

\subsection{SDF}
\begin{figure}[htbp]
  \includegraphics[width=1\linewidth]{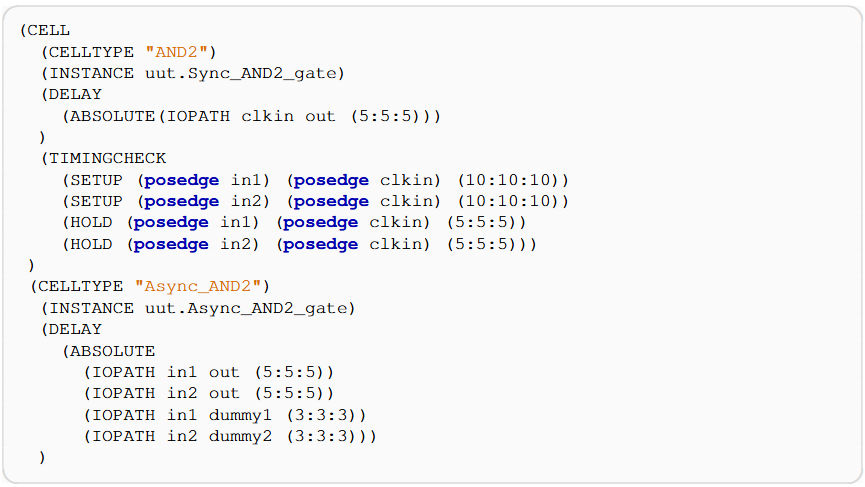}
    \vspace{-0.6cm}
\centering
  \caption{Example SDF}\label{fig:SDF}
\end{figure}

%Because specify blocks do not support hierarchical port references, we propose to use hierarchical references within the SDF file. 
For both synchronous and asynchronous designs, typical description of interconnect delay and IOPATH delay is possible. The pulse dissipation timing of asynchronous gates would be described as an IOPATH delay.  Fig.~\ref{fig:SDF}, shows example SDF delay and timing checks for synchronous and asynchronous gates. 

\section{Verification and Simulation}

\begin{figure}[htbp]
\includegraphics[width=1\linewidth]{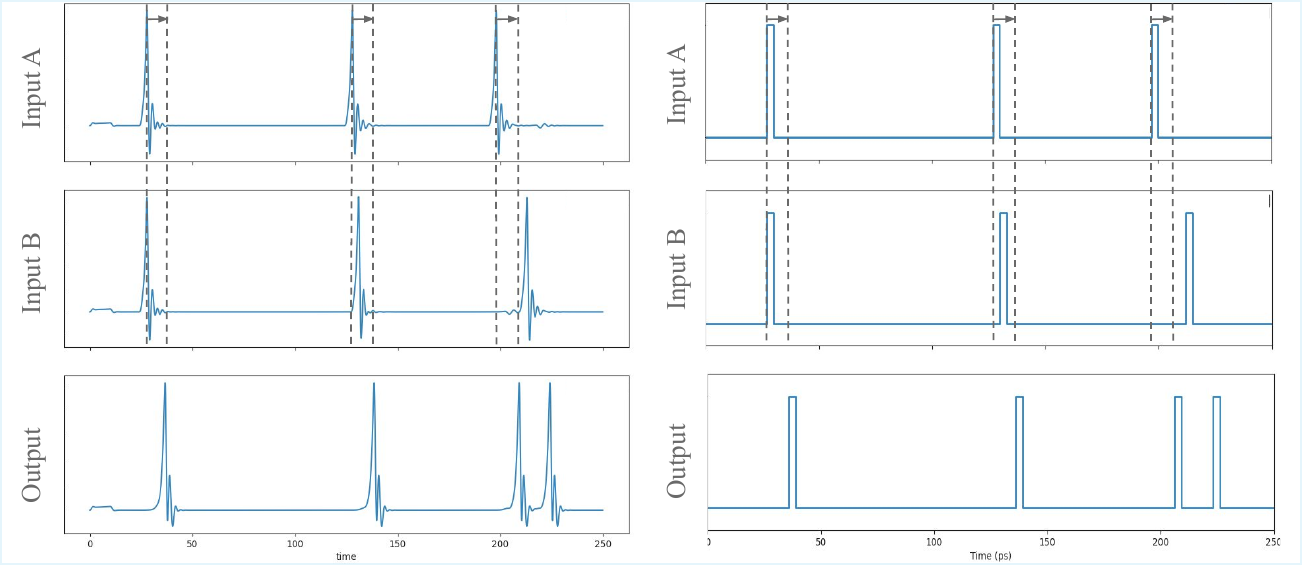}
    \vspace{-0.6cm}
\centering
  \caption{Josim and Verilog simulation of asynchronous OR gate}\label{fig:Sim_syncOR}
\end{figure}

JoSIM analog simulations were used to verify the behavioral and timing accuracy of our proposed Verilog model. This validation in particular contrasts the different application of retention windows for asynchronous gates. As demonstrated in Fig.~\ref{fig:Sim_syncOR}, asynchronous OR gates utilize retention windows as a pulse blocker rather than pulse persistence. The retention window prevents secondary input pulses from updating internal states, allowing only one output pulse generation in this time frame. Additionally, these validate proper clk-q delay through SDF back annotation.  

\begin{figure}[htbp]
  \includegraphics[width=1\linewidth]{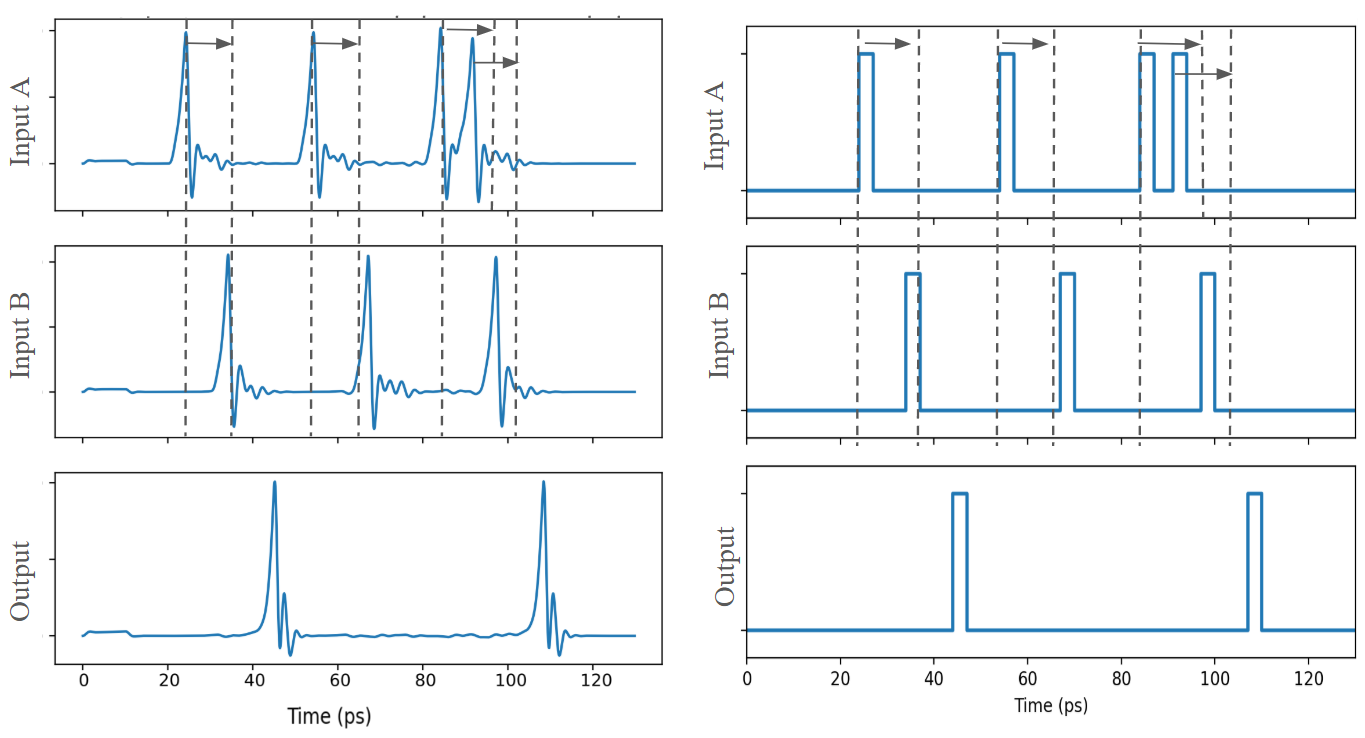}
    \vspace{-0.6cm}
\centering
  \caption{Josim and Verilog simulation of asynchronous AND with annotated retention windows}\label{fig:Sim_asynAND}
\end{figure}
In contrast to the blocking behavior of asynchronous OR gates, validation of asynchronous AND gates shown in Fig.~\ref{fig:Sim_asynAND} demonstrate output pulse generation under enforcement of coincidence within the retention window. Furthermore, the final generated retention window validates our model's support for window extension wherein a successive input pulse on the same port regenerated the retention window.

\begin{figure}[htbp]
  \includegraphics[width=1\linewidth]{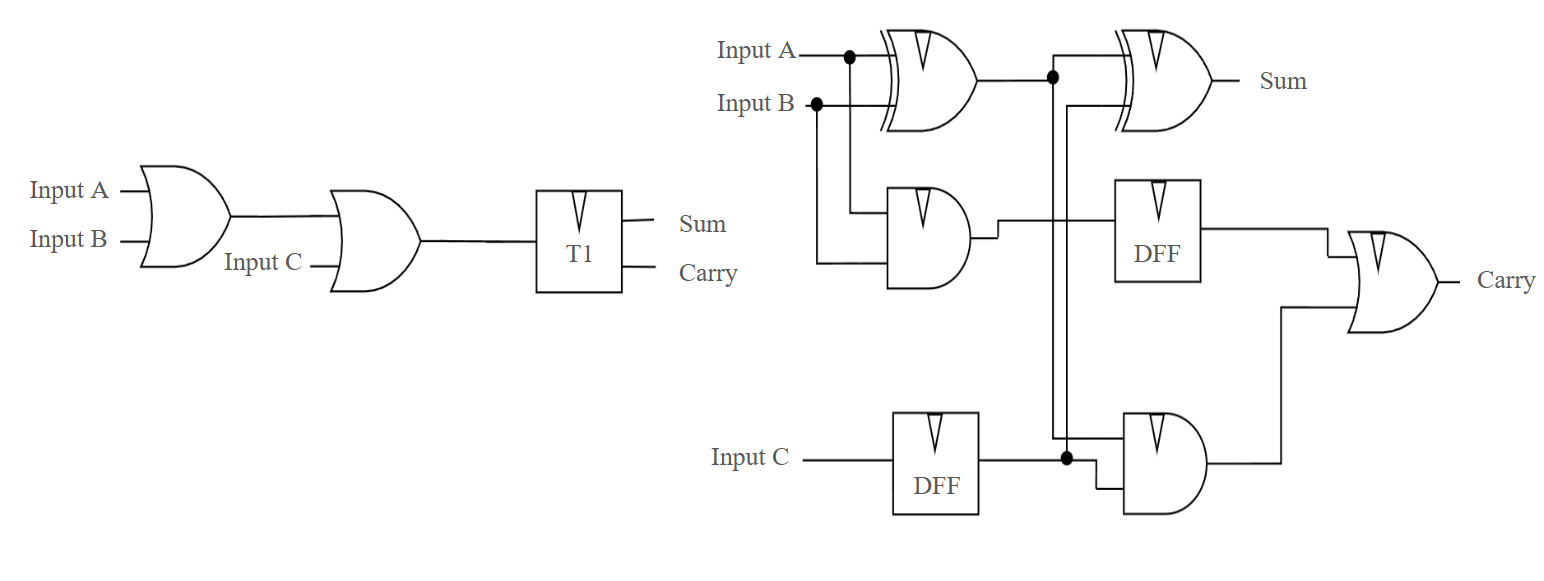}
    \vspace{-0.6cm}
\centering
  \caption{a) T1 cell FA  b) Synchronous design FA }\label{fig:1B_FA}
\end{figure}

\begin{figure}[htbp]
\centering
\includegraphics[width=\linewidth]{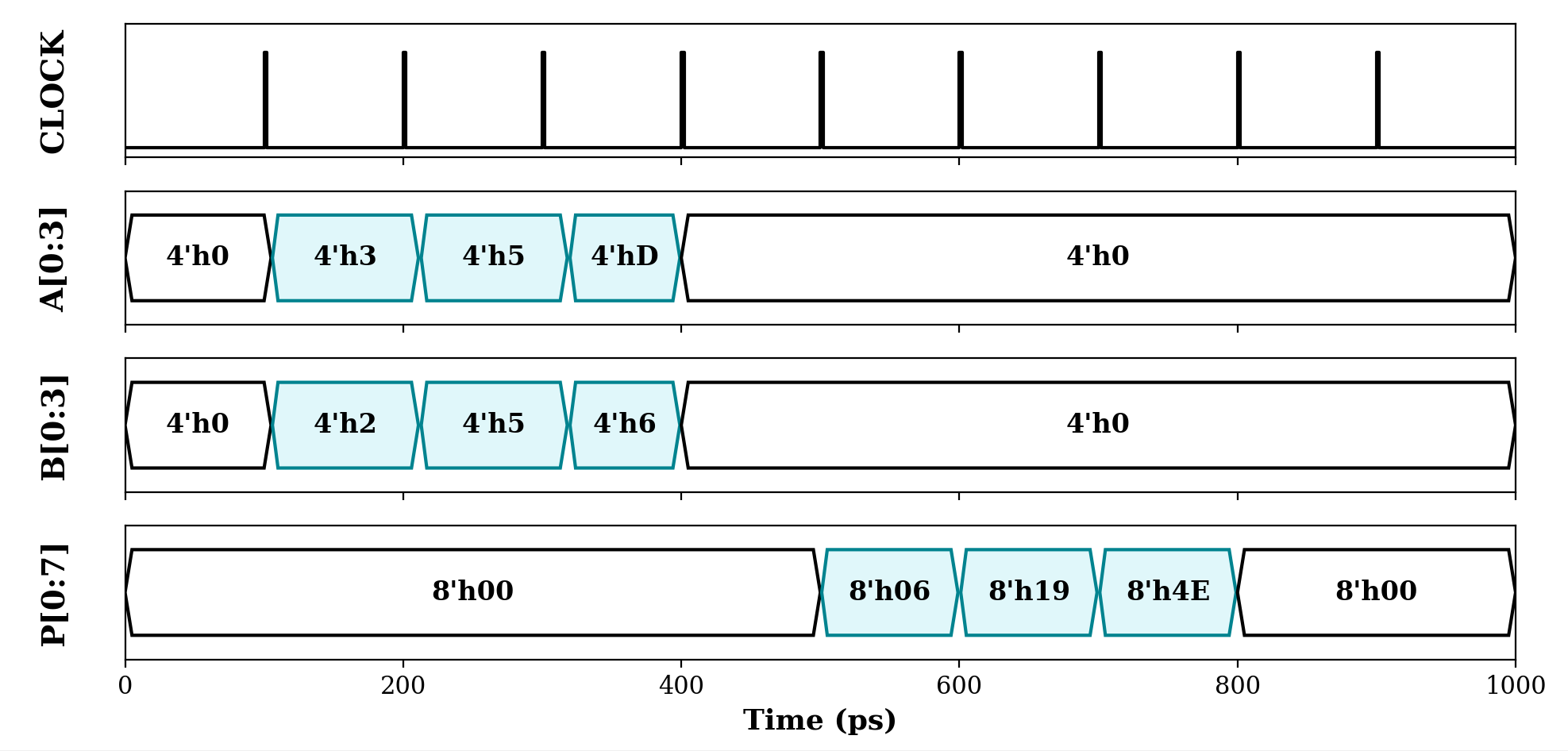}
\vspace{-0.6cm}
\caption{4-bit multiplier employing mixed gate usage}
\caption*{\footnotesize Note: For visualization, the waveform is shown using a sample-and-hold representation in which pulse values are held until the next clock cycle.}
\label{fig:4B_M}
\end{figure}

With this framework, circuits involving both synchronous and asynchronous gates can be explored and verified. As shown in Fig.~\ref{fig:1B_FA} a full adder implementation using two merger cells and a T1 cell reduces area consumption of a synchronous SFQ full adder approach within 1 clk cycle as opposed to 3 clk cycles, but introduces new timing constraints as input pulses must be timed to avoid collisions. A 4-bit multiplier can be constructed combining this full adder scheme with synchronous AND cells.  Using mixed asynchronous and synchronous gates a 4-bit multiplier can be created with 851 JJs as opposed to 4985 JJs if using only synchronous gates, highlighting the importance of supporting asynchronous gates in HDL.  In Fig.~\ref{fig:4B_M} we show waveform simulations of 4-bit multiplier consisting of both asynchronous and synchronous gates using our models.

\section{Conclusions}

As SFQ technologies promise high performance in ultra-low energy computing, adapting well-established CMOS EDA tools to support SFQ design flows will pave the way for continued research. Our proposed framework achieves accurate HDL abstraction for behavior and function simulation of circuit netlists, modeling both synchronous and asynchronous timing characteristics with full SDF back-annotation compatibility. Compared to previous approaches, our implementation reduces instantiation overhead while allowing greater design flexibility. Moreover, the structure of our proposed models enables straightforward extensions to other SFQ gate families such as self-resetting gates. This work represents an important step towards expanding EDA tool support for both current and future SFQ circuit design.

\bibliographystyle{IEEEtran}
\bibliography{references}
\end{document}